\documentclass[english,aps,pra,reprint,noshowpacs,longbibliography,showkeys]{revtex4-1}   
\usepackage[T1]{fontenc}	
\usepackage[latin9]{inputenc}	
\usepackage{geometry}		
\usepackage{graphicx}
\usepackage[above,below]{placeins}	
\usepackage{times}
\usepackage{hyperref}  
\hypersetup{colorlinks=true,urlcolor=blue,citecolor=blue,linkcolor=blue}
\begin{document}

\title{Network maps of student work with physics, other sciences, and math in an integrated science course}
\author{Jesper Bruun}
\affiliation{Department of Science Education, University of Copenhagen, Oester Voldgade 3, Copenhagen, Denmark}
\author{Ida Viola Andersen}
\affiliation{Niels Bohr Institute, University of Copenhagen, Blegdamsvej 17, 2100 Copenhagen OE, Denmark}


\begin{abstract}
In 2004 Denmark introduced a compulsory integrated science course the most popular upper secondary study program. One of the nation-wide course aims are for students to "achieve knowledge about some of the central scientific issues and their social, ethical, and historical perspectives". This is to be done via collaboration between the subjects, and often involves physics and another scientific subject. The official teaching plans further state that mathematics must be used for analysing data. We use network analysis to study six different implementations of the course in terms of the structure of different kinds of teaching/learning activities. By creating networks maps of each lesson, we show that teaching/learning activities in the course seldom tends to address how sciences can work together to solve a problem, but rather stages each natural science as a distinct and separate activity with a distinct identity.
\end{abstract}

\maketitle

\section{Introduction and background}
Cross-cutting skills and competencies are emphasized as important in many curricula across the world \citep{ananiadou200921st,national2012framework}. In the context of Science typical examples given of skills that the disciplines share are hypothesis generation and evaluation, creating and using models, designing and performing experiments and observation, and the ability to see Science in a broader perspective. One approach when teaching such skills is to use various forms of interdisciplinary teaching methods \citep{schaal2010concept}. However, interdisciplinary teaching is notoriously difficult to orchestrate\citep{ananiadou200921st}. It requires time, coordination, and a willingness for teachers to teach outside areas they would normally teach. And for some teachers, the level of the academic content might seem too low. These difficulties may lead to teaching-learning situations which may seem interdisciplinary on the surface, but are in reality just two or more disciplines working under the same very broad theme. 
\begin{figure}
\includegraphics[width=0.9\linewidth]{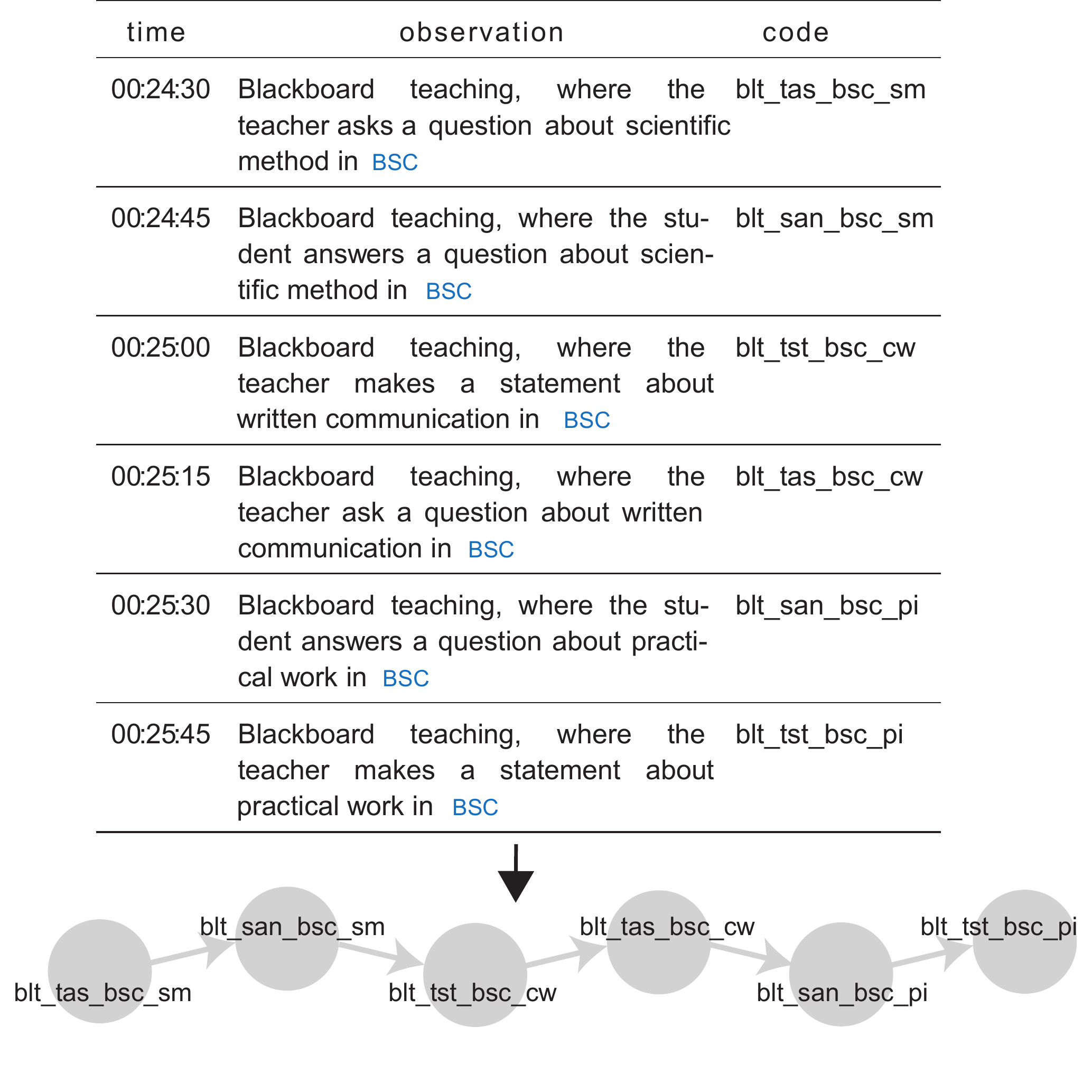}
\caption{\label{fig:networkcreation}Networks are created from 15-second interval codes that describe observed teaching.}
\end{figure}
As an analytical framework, Jantsch \citep{jantsch1947inter} developed four levels of interdisciplinarity: crossdisciplinarity (where questions are asked within the frame of a main discipline, and some technical help is provided by other disciplines), pluridisciplinarity (where disciplines work separately but in parallel under a common theme), interdisciplinarity (where disciplines are coordinated by a higher level problem or concept which cannot be addressed or properly explained by one discipline), and transdisciplinarity (where the problem takes center stage and the borders between disciplines may be blurred). Jantsch' framework can be used as a rough categorization, although his categories have been nuanced since then \citep{klein2010taxonomy}. In this paper, we use network analysis to map and characterize interdisciplinary teaching-learning situations both mathematically and visually. Our example is from a Danish course for upper secondary students. 

The Danish Basic Science Course (BSC) was introduced in 2004 as a compulsory integrated science course in the most popular of the country's upper secondary study program. One of the nation-wide course aims are for students to "achieve knowledge about some of the central scientific issues and their social, ethical, and historical perspectives". This is to be done via collaboration between the subjects, and often involves physics and another scientific subject. The learning aims of the course are aligned with the examples of concepts and competencies listed above. For example, students are to generate and test simple hypotheses, make observations and collect data, and to create and use simple models.

In our network description of course units, we create networks of codes that describe the activities, interactions, subjects, and learning goals for the BSC. See Figure \ref{fig:networkcreation}. We then partition the networks with a robust community detection algorithm called Infomap \citep{rosvall2008maps}. It is through this partitioning that we propose to relate network theory to Jantsch' conceptions of interdisciplinarity. We would expect pluridisciplinarity to be visible in a network as sharp divisions between different subjects, interdisciplinarity to be visible as a focus on general Science-related competencies, while transdiciplinarity would be visible as a mixing between subjects.  

We use two measures from network analysis\footnote{For details on these measures see Supporting Material (DOI: 10.13140/RG.2.2.12844.36486)} to analyze the resulting networks. The first measure is modularity, $Q$ \citep{newman2006modularity}. A network with sufficiently high modularity ($Q>0.3$,\citep{clauset2004finding}) consists of clusters (modules) of nodes that are tightly connected to each other but not to other modules. Thus, high modularity would signify teaching with sharp divisions between teaching activities. The second measure is the segregation, $D_{seg}$\footnote{Actually the Z-value ($Z=\frac{D-\overline{D}}{\sigma_D}$ over $n$ randomized versions of the network. We set $n=100$ for this study. See Supporting Material.}\citep{bruun2014time}, which which measures the tendency of node attributes to be overrepresented in modules. A network with perfect segregation on, say \emph{subject}, will be divided into modules with each subject appearing in separate modules. 

The point of employing these two measures is to characterize and to allow comparisons between teaching-learning situations. However,  Infomap not only creates modules but also uses uses transition probabilities between modules to calculate connections between modules. This results in a map of connected modules, which depicts over-arching trends in the network (\citep{bodin2012,lindahl2016,bruun2014time}). We investigate each module with regards to the labels used and the structure between labels. We use this to name each module and thus create a map of teaching. Since we use the same set of codes to describe each teachers' teaching, we can use diagrams that show how codes are distributed differently in different maps to compare teaching. Highly segregated modules, are easy to name, because the same code appears many times. We show an example of two maps from the same implementation (same class, different teachers) that illustrates how an image of interdisciplinarity emerges using this analysis. 

Our research questions are:

RQ1: What characterizes the six teachers' implementations when measured through modularity and segregation of codes that describe the subject addressed in observed and coded teaching-learning events?

RQ2: What images of disciplinary identity and interdisciplinary cooperation  emerge from this analysis of compartmentalization?

\section{Methods}
The data set consists of six networks of five course implementations with a total of 60 lessons of each 1.5 hours duration. Not all teaching in all implementations was observed, and except for implementation 3 (i3tC and i3tD in tables), only the physics teacher was observed. Table~\ref{tab:overview} gives an overview of the five implementations. All teachers except Teacher D taught physics. All implementations have a physics part and other scientific parts. Implement. In implementation 4, the physics teacher also taught earth science.  
\begin{center}
\begin{table}
\caption{\label{tab:overview}Overview of implementations. Different implementations are coded with an \emph{i} and a number. Different teachers are coded with a \emph{t} and a capital letter.}
\begin{tabular}{p{1.2cm}|p{1.5cm}|p{2cm}|c|c}
Teaching ID & Observed lessons & Subjects &Teacher &  Theme \\
\hline
i1tA & 15 of 32 &Phy. + Earth Sci. & A &  Myth Busters \\
i2tC & 13 of 33 &Phy. + Bio. & C & The Body \& Energy  \\
i3tC & 9 of 16 &Phy. + Bio. & C & The Body \& Energy  \\
i3tD & 8 of 13 &Phy. + Bio. & D &  The Body \& Energy  \\
i4tE & 11 of 33 &Phy.+Chem.+ Bio.+Earth Sci.&E & Global Warming\\
i5tF & 8 of 37 &Phy.+Chem. +Earth Sci.&F & Global Warming\\
\hline
\end{tabular}
\end{table}
\end{center}
Teaching was coded in 15 second intervals while observing and audio-recordings were used to verify codes. Codes in four distinct coding categories were developed to capture possible actions in the classroom. The categories were \emph{Activity} (mode of teaching, 15 different codes), \emph{Interaction} (dialogical actions made by teacher and students, 11 different codes) \emph{Subject} (8 codes: Science (as in the general aims for the BSC course), Physics, Biology, Innovation, Mathematics, Chemistry,  Earth Science, Non-academic content), and \emph{Learning Aim} (skills that the activity is focusing on, 8 codes). Thus, each 15-second interval was labeled with a four-tier code (see Figure~\ref{fig:networkcreation}). Networks were then created based time ordering of labels. An arrow of thickness one is drawn from the node labeled \textsc{blt\_tas\_bsc\_sm} to the node labeled  \textsc{blt\_san\_bsc\_sm}. The activity in both cases was blackboard teaching (\textsc{blt}), the subject addressed lay within the basic science course (\textsc{bsc}), and the event was about scientific methods (\textsc{sm}). The difference between the two events were that first the teacher asked a question (\textsc{tas}) and then the student answered that question (\textsc{san}). The network can be described by a weighted adjacency matrix, A. For each time label $i$ follows four-tier code $j$, one is added to the matrix element $a_{ij}$. In this way networks of each teacher's lessons were created. 

We used Infomap to partition each network and tested the robustness of our results \citep{breweBruun2016}. We ran Infomap 1000 times for each course implementation to gauge the variability in Infomaps community detection. For each partition found, we calculated the modularity, $Q$, $Z_D$ with respect subjects addressed, and the \emph{Normalized Mutual Information} (NMI, see e.g. \citep{lancichinetti2009community}) of the given partition and the 999 other partitions. The NMI is a standard measure used in community detection to measure how alike two partitions are. NMI ranges from 0 to 1, with 1 meaning a perfect overlap. We report the average values of each of these measures. For each teacher's teaching, we also calculated the frequency with which subject codes were observed. 

Finding that Infomap consistently reproduces almost the same modules for all networks (see below), we could safely pick two solutions for our illustration. We generated maps and an alluvial diagram \citep{bohlin2014community} of codes for these solutions. We then named modules according to the prevalence and structure of the codes that appeared in each module. 

\section{Results}
\begin{center}
\begin{table}
\caption{\label{tab:solution} Number of nodes ($N$), unique links ($N_l$), mean modularity $\overline{Q}$ , mean NMI, and the mean segregation Z-value, $\overline{Z}_D$. Numbers in parenthesis are the standard errors on the last digit.}
\begin{tabular}{c|c|c|c|c|c}
Network & $N$ & $N_l$ &$\overline{Q}$ & $\overline{NMI}$ & $\overline{Z}_D$ \\
\hline
i1tA & 105 &395 &0.7571(1) & 0.9750(2) & 14.03(1) \\
i2tC & 109 &384 &0.6761(0) & 0.9457(5) & 13.31(1) \\
i3tC & 90 &354 &0.5629 & 1 & 13.78(1) \\
i3tD & 87 &292 &0.7262(0) & 0.9960(1) & 13.88(1) \\
i4tE & 115 &453 &0.8011(0) & 0.9995(0) & 19.57(1) \\
i5tF & 113 &370 &0.7392(2) & 0.9584(3) & 9.56(2) \\
\hline
\end{tabular}
\end{table}
\end{center}
\begin{figure}
\includegraphics[width=0.9\linewidth]{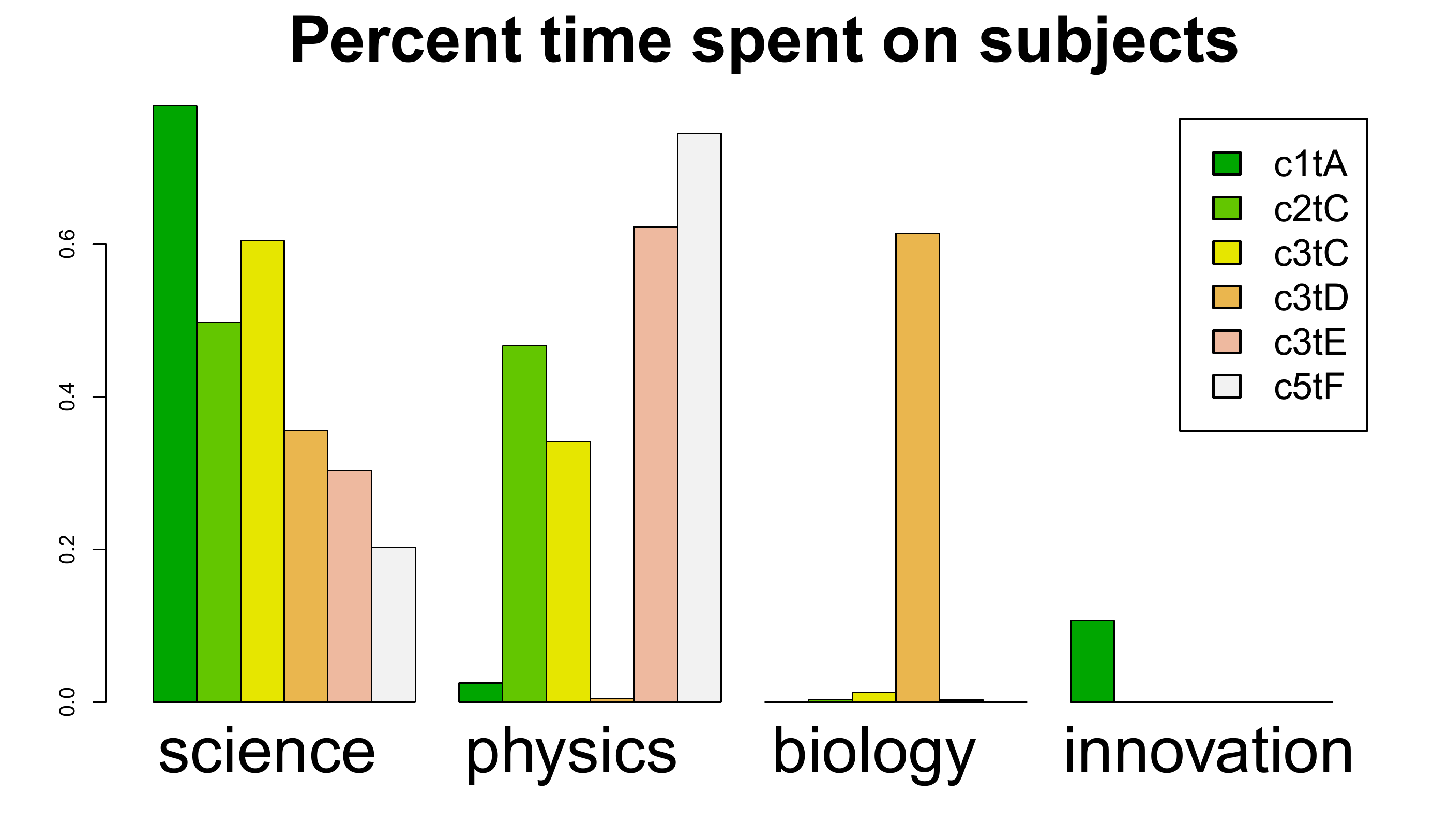}
\caption{\label{fig:timeSpentSubject} The percent of time spent addressing different subjects in each teacher's classroom.}
\end{figure}
The observations of the six teachers resulted in six networks. Table~\ref{tab:solution} shows the results of our calculations along with the number nodes (unique labels) and unique links (connections between labels). All networks are characterized by high modularities ($\overline{Q}>0.5$), which is comfortably above the 0.3 threshold for modularity. Thus, they are all modular (many of them very much so). Infomap is very consistent ($\overline{NMI}>0.94$) in partioning the networks. Finally, the Segregation Z-values are very high, which means that individual subjects are addressed in distinct and separate modules and not in a way, for example, that alternates between subjects. 

Figure~\ref{fig:timeSpentSubject} shows how much of the time for each teacher was spent on addressing science as a discipline in its own right, and how much was spent on the individual disciplines. Teachers vary a lot, but interestingly, in Implementation 3, the Physics teacher (tC) splits the time roughly equally between Science and Physics, while the Biology teacher most of the time addresses Biology. Also, teacher D's network has a higher modularity than teacher C's for the same implementation. They have comparable segregation, number of nodes and links, and they taught the same implementation, so it is interesting to compare these two teachers in terms of a more in-depth analysis as described above. 

This analysis shows that Teacher C's teaching is characterized by a central module where C lectures about all aspects of science covered in the BSC curriculum.  C seems to use this kind of lecturing as a backdrop for talking about physics as a discipline, and about the scientific method as such. Apart from the lecture style, much of the teaching is of the form Questions \& Answers, where students ask questions and the teacher answers. In contrast Teacher D focuses on asking questions to students about that pertains to Biology as a discipline, and uses questions to stage Biology as a scientific discipline. The latter is done by lecturing and questioning about how Biology relates to the specific aims described in the BSC curriculum. 

\begin{figure*}
\includegraphics[width=0.9\linewidth]{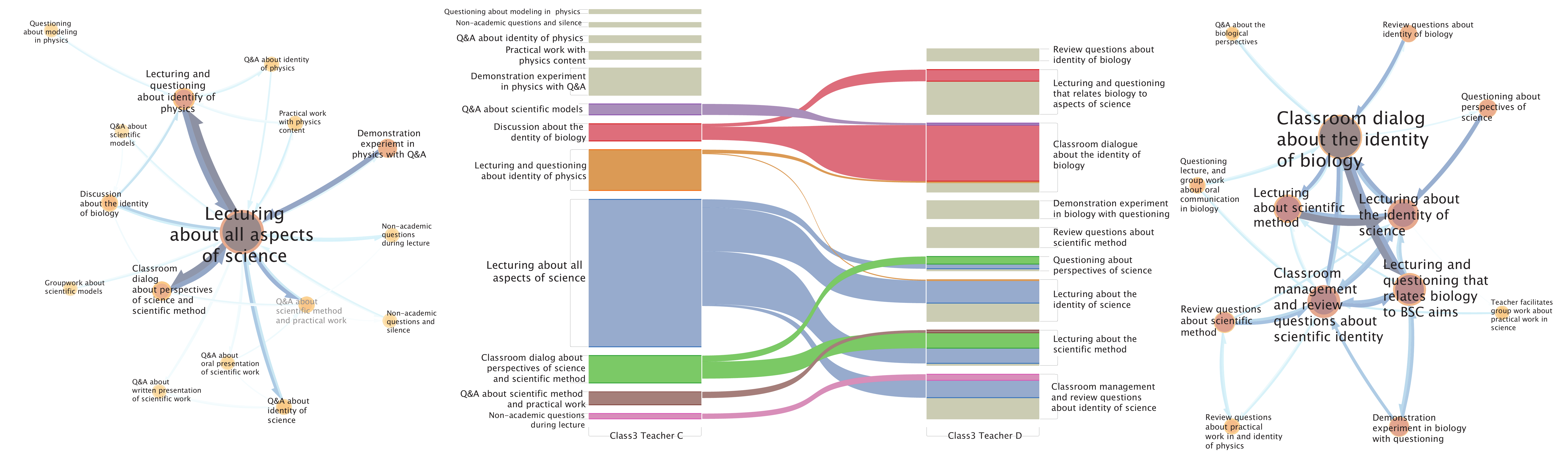}
\caption{\label{fig:comparison}The left-hand map shows the map for Teacher C. The right-hand map shows the same for Teacher D. Each module (circle) has internal structure between labels. Links between modules represent teaching-learning activities in one module being followed by activities in another. Text on modules represent our interpretation of the internal structure in a module. The stream lines in the alluvial diagram in the middle shows overlap in labels used to describe observed teaching. For instance, the blue rectangle is split into four different modules, representing that Teacher C addressed all aspects of science in a unified manner, while Teacher D had separate activities for different aspects. Chunks of rectangles without connecting streamlines resemble codes that are not shared between the two teachers' teaching.}
\end{figure*}

\section{Discussion}
We argue that both teacher C and D focus on staging their discipline as an example of a scientific discipline. They share a common theme, which is the Body and Energy, but we see no signs of cooperation between subjects. The questions asked by the teachers do not transcend the boundaries of their respective disciplines. Thus, Jantsch's category pluridisciplinarity seems to describe this teaching most aptly. 

The two teachers' lessons were intertwined in the sense that one teacher would teach some modules and in between those the other teacher would teach. Given the teachers' different ways of teaching and their focus on staging their own discipline, one could argue that students would experience only a fragmented link between Physics and Biology. They may see that the body and energy is a theme that can be addressed by both Physics and Biology, but may not become aware of any synergy effects between these to disciplines. This is in contrast to many real world problems, which require that disciplines work together across their respective boundaries. 

Given the high segregation, modularities and much time spent on individual disciplines, most of the investigated implementations see to be pluridisciplinary. However, Implementation 1 (Myth Busters) is different. Here, most of the time is used on activities pertaining to Science as a general discipline. A small percentage of the time is used on innovation (which we have included here as a separate discipline, since a particular innnovation model was taught), but the subjects are still very segregated. This could be an example of Jantsch's \emph{interdisciplinarity}, in which disciplines are subjected to a common higher level concept; in Implementation 1 the particular disciplines would only be addressed if the myths, which students wanted to bust, required them. 

\section{Conclusion}
We transformed observations of five implementations of the Danish BSC-course into networks. We then used Infomap to partition the networks into modules and calculated modularity, segregation and time spent on disciplines for each of the implementations. Implementations were highly modular and highly segregated. We took this to signify that implementations were pluridisciplinary meaning that subjects worked mostly in parallel to each other. A deeper analysis of one implementations showed two teachers who used the BSC as a stage for discussing the identity of a particular discipline in relation to the more general area of Science. 

\bibliography{Bruun_PERC}  	
\end{document}